\RequirePackage{fix-cm}

\documentclass[twocolumn]{svjour3}          
\smartqed  
\usepackage{graphicx}
\usepackage{amsmath}
\usepackage{color}
 \journalname{Granular Matter}

\newcommand{\edits}[1]{{\color{black}{#1}}}
 
\begin{document}
\title{Edges control clustering in levitated granular matter}

\author{Melody X. Lim        \and
		Kieran A. Murphy	 \and
		Heinrich M. Jaeger
}

\institute{M. X. Lim \and
		K. A. Murphy\and
		H. M. Jaeger\\
              Department of Physics and James Franck Institute \\
              The University of Chicago \\
              Chicago, Illinois 60637 USA \\
              \email{jaeger@uchicago.edu}    
}

\date{Draft: \today}

\maketitle

\begin{abstract}
The properties of small clusters depend dramatically on the interactions between their constituent particles. However, it remains challenging to design and tune the interactions between macroscopic particles, such as in a granular material.  Here, we use acoustic levitation to trap macroscopic grains and induce forces between them. Our main results show that particles levitated in an acoustic field prefer to make contact along sharp edges. The radius of curvature of the edges directly controls the magnitude of these forces. These highly directional interactions, combined with local contact forces, give rise to a diverse array of cluster shapes. Our results open up new possibilities for the design of specific forces between macroscopic particles, directing their assembly, and actuating their motion.

\end{abstract}

\section{Introduction}

Shape provides an important means by which to tune key properties of dense aggregates, whether their packing fraction~\cite{Donev2004,Degraaf2011,Roth2016}, contact configuration~\cite{Haji2009,Damasceno2012,Neudecker2013,Behringer2014,Walsh2016}, or mechanical properties~\cite{Miszta2011,Azema2012,Miskin2013,Farhadi2014,Zhao2016,Tang2016,Murphy2018}. In each of these cases, the shape of a particle determines the global properties of an aggregate by modifying \edits{both the geometry of the packing and} local inter-particle contacts. These local contacts play a particularly important role in jamming, where the precise type of contact between faceted shapes may dramatically change the mechanical stability of a given packing. In addition, breaking spherical symmetry in a dense granular packing produces a coupling between normal forces and torques on the single-grain level.  

In addition to local contact forces \edits{and geometry}, which produce effects at the level of a single contact on a grain, longer-range particle interactions also play an important role in the properties of a granular material. Previous work has shown that adding weakly attractive dipole-dipole interactions to a granular material provides a means of self-organisation for the force chains that form under compression, thus strengthening the force network that forms as a result~\cite{Cox2016}. Alternatively, longer range forces can drive clustering~\cite{Royer2009,Lee2015} and particle motion~\cite{Lumay2008,Cavallaro2011}. However, the relation between attractive forces, particle shape, structure, and mechanical properties remains to be fully understood. 

In this article, we take a first step towards this goal by producing shape-dependent, tunable attractive forces between granular materials using acoustic levitation. Previous work on acoustically levitated particles has focused on the force between levitated particles with a high degree of symmetry, such as spheres~\cite{Gorkov1961,Doinikov1994,Bruus2012,Settnes2012,Silva2014}, \edits{cylinders~\cite{Wei2004,Wang2011,Mitri2016}}, or between smooth deformable objects such as liquid drops~\cite{Crum1971} and bubbles~\cite{Leighton1990}. \edits{Other work has focused on viscous and thermal effects in the acoustic force~\cite{Westervelt1951,Doinikov1994,Doinikov1997}. More recently, finite-element simulations have been shown to be an accurate, general tool for the calculation of acoustic radiation forces for particles of arbitrary shape~\cite{Glynne2013,Hahn2015}, for a variety of boundary conditions~\cite{Leibacher2015}.} 

We present results for a wide range of particle shapes and constituent materials, showing that particles are attracted to one another along sharp edges. Using finite element simulations, we demonstrate that the strength of the inter-particle attraction is controlled by the sharpness of particle features. These bonds also support robust hinge-like motions, which we suggest are oscillations about a minimum in some energetic landscape.  Finally, we consider the \edits{structure of} clusters formed in the acoustic field. We suggest that the unique structures we observe are due to a combination of the highly anisotropic acoustic forces, packing efficiency, and stabilisation from local inter-particle contacts. 

\section{Methods}

\begin{figure}
\includegraphics[width=\columnwidth]{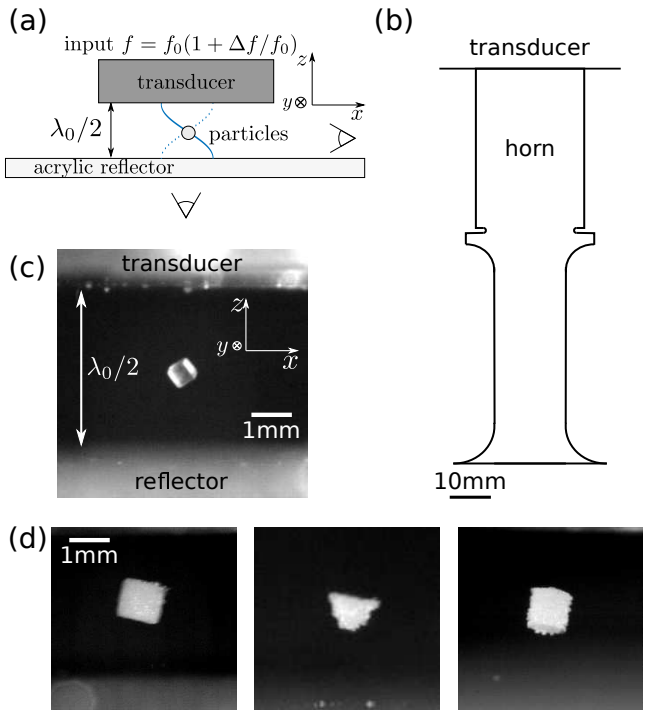}
\caption{(a) A schematic of the apparatus. An ultrasound transducer generates sound waves in air (speed of sound $c_s=343m/s$). \edits{The transducer consists of a commercial ultrasound transducer, the output of which is amplified by an aluminium horn bolted onto its base.} The distance between the \edits{base of the horn} and a transparent acrylic reflector is chosen so as to create a pressure standing wave with a single node, with frequency~$f_0=45.65kHz$ and wavelength~$\lambda_0=c_s/f_0$. Particles are levitated in this pressure node, and can be imaged from the side or from below. The driving frequency~$f$ of the trap can be detuned by~$\Delta f \equiv f-f_0$, inducing active fluctuations in the levitated particles~\cite{Rudnick1990,Lim2018}. (b) \edits{A scale drawing of a cross-section of the combined transducer and horn (which are circularly symmetric). Particles are levitated beneath the bottom surface of the horn.} (c) A grain of table salt levitated in the acoustic trap, imaged from the side. (d) 3D printed particles, levitated in the acoustic trap. From left to right: cube (1mm side length), tetrahedron (1mm side length), cylinder (1mm diameter, 1mm height)}
\label{fig:setup}    
\end{figure}

We levitate grains using the setup illustrated in Fig.~\ref{fig:setup}(a). An ultrasound transducer (Hesentec \edits{HS-4AH-3540, base diameter 55mm, frequency 40$\pm$1 kHz}) generates ultrasound, which is amplified by an aluminium horn bolted onto the transducer (base diameter 38mm). \edits{A schematic of the horn, which is based on the design reported in Ref.~\cite{Andrade10}, is shown in Fig.~\ref{fig:setup}(b). In order to find the resonant frequency of the transducer and horn, we invert the setup (so the bottom surface of the horn is facing upwards) and scatter 100$\mu$m polyethylene grains on the surface. We then adjust the driving frequency slowly until we find the peak displacement of the grains on the transducer surface. This defines the resonant frequency of the transducer (coupled to the horn), $f_0=45.65$kHz.} The transducer is then driven at frequency~$f_0$ by applying an AC signal with peak-to-peak voltage 200V. 

A transparent acrylic reflector \edits{($l \times w \times h=152\times 152\times 6$mm$^3$)} is \edits{glued to a box, which is} mounted to a lab jack and adjusted so that the distance between transducer and reflector is~$\lambda_0/2$, corresponding to \edits{the transducer} resonant frequency~$f_0$, \edits{or wavelength~$\lambda_0 \approx 7.2$mm in air}. \edits{The setup (including transducer) is enclosed in an acrylic box ($l \times w \times h=0.6\times 0.3\times 0.6$m$^3$) to mitigate the effect of side-wind perturbations. The walls of the box are far from the levitation area, such that the acoustic trap has open boundary conditions.} Particles levitate in this standing wave, within a horizontal plane halfway between the reflector and transducer. Levitated grains can be imaged from the side or from the bottom. As shown in previous work~\cite{Lim2018}, \edits{driving the acoustic trap with frequency slightly above resonance} induces active fluctuations in the levitated particles. In turn, these active fluctuations drive cluster rearrangements. \edits{In this case, we induce active fluctuations by detuning the trap to~$f=45.76$kHz, or~$\Delta f/f_0=2.5\times 10^{-3}$. We note that within the range of frequencies used for detuning, there is no noticeable change in the displacement characteristics of the transducer. }

A wide range of particle shapes and constituent materials can be levitated in this setup. We take advantage of the naturally cubic shape of salt grains (material density 2 160 kg m$^{-3}$) to levitate isotropic cubes (Fig.~\ref{fig:setup}(c)). Alternatively, we levitate spheres (polyethylene, diameter 710-850$\mu$m, Cospheric, material density 1 000 kg m$^{-3}$). Finally, in order to create particles with controlled shapes and sizes, we 3D print cubes (1mm side length), tetrahedra (1mm side length), cones (1mm base diameter, opening angle 60$^\circ$), and cylinders (diameter 1mm, height 1mm). All particles are printed in hard UV-cured plastic (Objet VeroWhite, material density 1 180 kg m$^{-3}$), on an Objet Connex 350 printer. Note that the particles are printed in stacked laminae (thickness $\sim$50$\mu$m), producing highly frictional faces (Fig.~\ref{fig:setup}(d)).

\section{Results}

A single particle in the acoustic trap rotates freely about its centre of mass, \edits{regardless of shape and material}, confined by the \edits{primary sound field} in the vertical direction. When a second particle is introduced, \edits{the primary sound field confines it to the same levitation plane as the first particle. In addition, acoustic scattering due to the presence of the first particle produces an additional in-plane force, which we refer to as the secondary acoustic force.} This \edits{secondary (induced)} acoustic force, \edits{which for deformable particles is known as the secondary Bjerknes force~\cite{Garcia2014,Silva2014,Mohapatra2018}}, stabilises compact clusters. \edits{For spherical (isotropic) particles, the acoustic force follows the axisymmetry of the particles in the levitation plane, such that spherical particles cluster into a close-packed, two-dimensional lattice~\cite{Lim2018}}. However, since the acoustic force is generated by scattering, anisotropy in the particle shape results in anisotropy in the acoustic force. 

\begin{figure}
\includegraphics[width=\columnwidth]{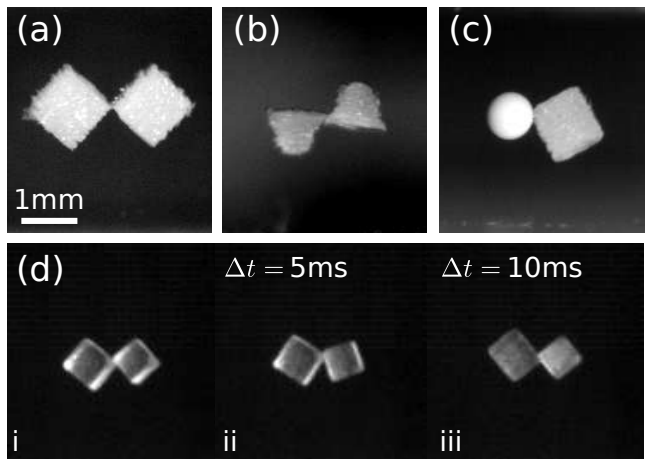}
\caption{Pairs of levitated particles imaged from the side. All images are shown to the same scale. (a) Two 3D printed cubes attach via a single edge. (b) Two cones attach by sharing an edge. We also observe configurations where both cones are pointed upwards or downwards. (c) A polyethylene sphere (800$\mu$m) attaches to the edge of a cylinder. (d) Sequence of images from the side showing a pair of salt grains executing a hinge motion. See Supplementary Movies for dynamics}
\label{fig:pairwise} 
\end{figure}

Due to this anisotropic force, \edits{when a second cube is introduced to the acoustic trap, the freely rotating cube is stabilised:} a pair of levitated cubes pack by sharing a single edge, with the centre of masses of both cubes in the nodal plane, as shown in Fig.~\ref{fig:pairwise}(a).  This configuration contrasts sharply with the arrangements that are generated through dipole-dipole interactions~\cite{Vutukuri2014}, depletion forces~\cite{Rossi2011}, or entropic considerations~\cite{Zhao2011,vanAnders2013,VanAnders2014}. In order to confirm that the cluster configuration we observe are driven by anisotropic acoustic forces, we levitate different combinations of shape pairs. Two cones also cluster by aligning a single edge, with the base of both cones in the nodal plane (Fig.~\ref{fig:pairwise}(b)). In addition, levitating a cylinder with a sphere produces a cluster where the sphere attaches to the sharp edge of the cylinder (Fig.~\ref{fig:pairwise}(c)). 

These edge-mediated configurations are stable to fluctuations induced by detuning the trap (see Supplementary Movies). Perturbing the cluster produces periodic variations in the contact angle between the cubes, as in Fig.~\ref{fig:pairwise}(d). These hinge-like motions suggest the existence of a restoring force towards the equilibrium configuration shown in Fig.~\ref{fig:pairwise}(a). Increasing the amplitude of active fluctuations by detuning the acoustic field results in the cluster breaking due to energetic collisions with the reflector. We do not observe face-face contacts when the constituent parts of the cluster reassemble. This observation, combined with the existence of periodic hinging motions, suggests that these edge-edge contacts are to be understood as the minimum of some energetic landscape.  

In order to shed light on the energetic landscape underlying the stability of edge-edge contacts, we calculate the acoustic potential around a perfectly scattering cube using finite-element simulations.  This acoustic potential \edits{is generated by acoustic scattering from the levitated particles, and can be calculated via a perturbation expansion of the acoustic fields in the levitation medium~\cite{Gorkov1961,Bruus2012}. Within this approximation,} the acoustic potential~$U_\mathrm{rad}$ on a scatterer with radius~$a$ (much smaller than~$\lambda_0$), speed of sound~$c_p$, and material density~$\rho_p$ in an inviscid fluid with speed of sound~$c_0$ and density~$\rho_0$ is

\begin{align}
U&=\frac{4\pi}{3}a^3 \rho_0\left[f_1\frac{1}{2}c_0^2 \langle p^2\rangle - f_2\frac{3}{4}\langle v^2\rangle\right],
\label{eq:Urad}
\end{align}
where angled brackets denote time averages of the pressure~$p$ and velocity~$v$. The scattering coefficients~$f_1$ and~$f_2$ are given by
\begin{align*}
f_1&=1-\frac{c_p^2\rho_p}{c_0^2\rho_0}\\
f_2&=\frac{2(\rho_p/\rho_0-1)}{2\rho_p/\rho_0+1}.
\end{align*}

\edits{We emphasise that Eq.~\ref{eq:Urad} is quantitatively accurate only for spherical (point-like) scatterers, in the Rayleigh limit ($a\ll \lambda_0$). However, our particles are closer to the Mie regime ($a \approx 0.1\lambda_0$ in this case), and in addition have finite extent. Our simulation results should thus be interpreted as an approximation of the true acoustic forces on our experimental particles. Nevertheless, they provide a qualitative demonstration of the acoustic forces on a point object due to the presence of a single object with sharp edges, and provide a useful starting point to explain our experimental observations. Future work could focus on a full simulation of the interaction between a pair of cubes.}

We use finite-element simulations (\edits{COMSOL}) to calculate the acoustic potential on a point scatterer in the vicinity of a single cube. In these simulations, we reproduce the experimental conditions (in the frequency domain) by driving the upper boundary at constant normal acceleration, then establishing perfectly reflecting boundary conditions on a parallel surface half a wavelength below. \edits{The simulation domain contains only the levitation volume (we do not simulate outside of the boundaries of the transducer).} Plane-wave radiation conditions on the lateral boundaries dissipate the acoustic field.  Within this volume, we fix a perfectly scattering cube with side length~$l=0.1\lambda_0$. Given these boundary conditions, we compute the time-averaged quantities~$\langle p^2\rangle$ and~$\langle v^2\rangle$ within the geometry of the trap. \edits{The acoustic fields are computed by resolving the acoustic wave equations on a physics-controlled mesh, with element size set to ``extremely fine" (maximum element size $2.94\times 10^{-4}$, minimum element size $5.88\times 10^{-7}$, maximum element growth rate 1.1).}

\begin{figure}
\includegraphics[width=\columnwidth]{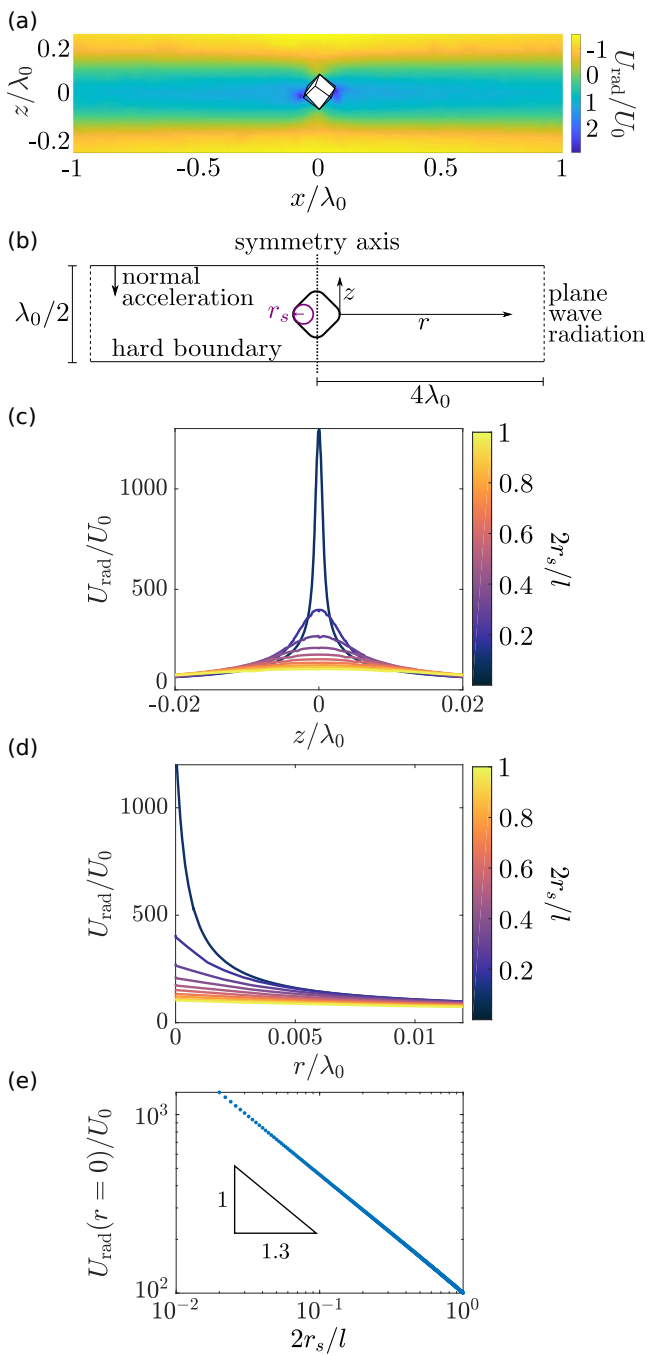}
\caption{(a) Cross-section of the simulated acoustic potential~$U_\mathrm{rad}$ around a perfectly scattering cube (cross-section in white, drawn schematically using black lines) in an acoustic standing wave. We normalise~$U_\mathrm{rad}$ by~$U_0$, the primary acoustic potential on a point particle in the center of the trap. (b) Schematic of simplified axisymmetric simulation geometry. The acoustic potential is calculated around the surface of revolution of a rounded square with side length~$l$ and corner radius of curvature~$r_s$ and side length~$l$, fixed in the center of a standing wave. (c)~$U_\mathrm{rad}/U_0$ as a function of vertical distance from the levitation node~$z$, for different~$r_s/l$. (d)~$U_\mathrm{rad}/U_0$ as a function of radial distance from the particle surface~$r$, for different~$r_s/l$. (e) Depth of potential minimum,~$U_\mathrm{rad}(r=0)/U_0$, as a function of~$r_s/l$ }
\label{fig:comsol}    
\end{figure}

Substituting the simulated pressure and velocity fields into Eq.~\ref{eq:Urad} thus produces the acoustic potential around a perfectly scattering cube, acting on a point scatterer with radius~$a=0.1\lambda_0$, speed of sound~$c_p=$2 620 m s$^{-1}$, and density~$\rho_0=$1 180 kg m$^{-3}$ (material parameters were chosen to match the \edits{3D-printed} experimental grains). \edits{In order to most clearly show the contribution of the edges, we orient the cube such that its longest diagonal is in the~$z$-direction.} A cross-section of the three-dimensional acoustic potential is shown in Fig.~\ref{fig:comsol}(a). The acoustic field shows a strong gradient in the~$z$ direction, due to the primary confining acoustic force -- a levitated particle minimises energy by having its centre of mass in the nodal plane ($z=0$). In addition to the primary confining acoustic field, secondary acoustic scattering results in potential minima located at regions of high curvature (the edges of the cube), indicating that a point particle in the acoustic trap would be disproportionally attracted to the edges of the cube. 

The boundary conditions at the surface of a levitated particle connect geometry and acoustic forces. In particular, we model a perfectly scattering surface using the Neumann boundary condition. For an acoustic wave incident on a perfectly scattering surface with unit normal~$\hat{n}$, in an inviscid fluid with pressure~$p$, the boundary condition reads

\begin{align}
    \hat{n}\cdot \nabla p=0.
    \label{eq:bc}
\end{align}
Eq.~\ref{eq:bc} implies that the pressure gradient at a perfectly scattering surface must be orthogonal to the surface normal. In this case, much like a conducting surface in an electric field, the pressure gradient is forced to change with the curvature of the surface. We suggest that this effect focuses the acoustic field in regions of high curvature, thus producing deep potential minima at the edges of the cube.

We illuminate the relation between geometry and acoustic force by quantifying the effect of a single edge on the acoustic potential.  Instead of simulating the acoustic potential around a fully three-dimensional cube, we instead use an axisymmetric geometry (boundary conditions shown in Fig.~\ref{fig:comsol}(b)), and calculate the acoustic field around the surface of revolution of a square with side length~$l=0.1\lambda_0$. We parameterise the edge sharpness of the square using the radius of curvature~$r_s$ of the corners. Our results for the acoustic potential in the~$z$- and~$r$-directions as a function of nondimensional edge curvature~$r_s/l$ are shown in Figs.~\ref{fig:comsol}(c) and (d). As~$r_s/l$ decreases from 1 (sphere) to 0 (sharp corners), the depth of the acoustic potential increases sharply in both the~$z$- and~$r$-direction.

\edits{In order to further quantify the relation between edge sharpness and acoustic force, we simulate the normalised depth of the potential minimum at the surface of the particle,~$U_\mathrm{rad}(r=0)/U_0$, as a function of~$r_s/l$. Data is plotted in Fig.~\ref{fig:comsol}(e), and reveal that the depth of the acoustic potential diverges as a power law as the edge curvature approaches zero, with exponent~$-10/13$.} Our results confirm that the near-field force between a point scatterer and a levitated particle in an acoustic field scales sensitively with the sharpness of the edges on the levitated particle. 

Based on our simulations of the acoustic field, we suggest an explanation for the robustness of edge-edge contacts and hinge-like motions between a pair of levitated cubes. When two cubes are levitated, their centres of mass are confined to the nodal plane due to the primary acoustic field. In addition, the secondary acoustic force produces short-range attractive forces, which preferentially align the edges of faceted shapes. Upon joining a pair of cubes in the configuration shown in Fig.~\ref{fig:pairwise}(a), we note that rotation of a two-cube cluster around their common edge results in motion of the centres of mass out of the nodal plane. The hinging motion shown in Fig.~\ref{fig:pairwise}(d) thus arises from the interplay of geometry-controlled interactions and the confining acoustic field. 

The geometry of the cubes encourages packing in a square lattice at high packing fractions. At the same time, however,~\edits{in an acoustic trap,} the edges of the cubes give rise to interactions that favour edge-edge contacts \edits{(such as illustrated in Fig.~\ref{fig:pairwise}(a))}. Since each cube can only make two edge-edge contacts~\edits{(one on each side of a cube positioned as in Fig.~\ref{fig:pairwise}(a))}, as the cluster size grows, it becomes harder and harder for all cubes in the cluster to satisfy their constraints. The acoustic and contact forces that drive self-assembly in an acoustic trap thus also produce highly frustrated clusters. Such clusters exhibit a multiplicity of particle arrangements, transitions between which require extensive reconfigurations.  

\begin{figure*}
\includegraphics[width=2\columnwidth]{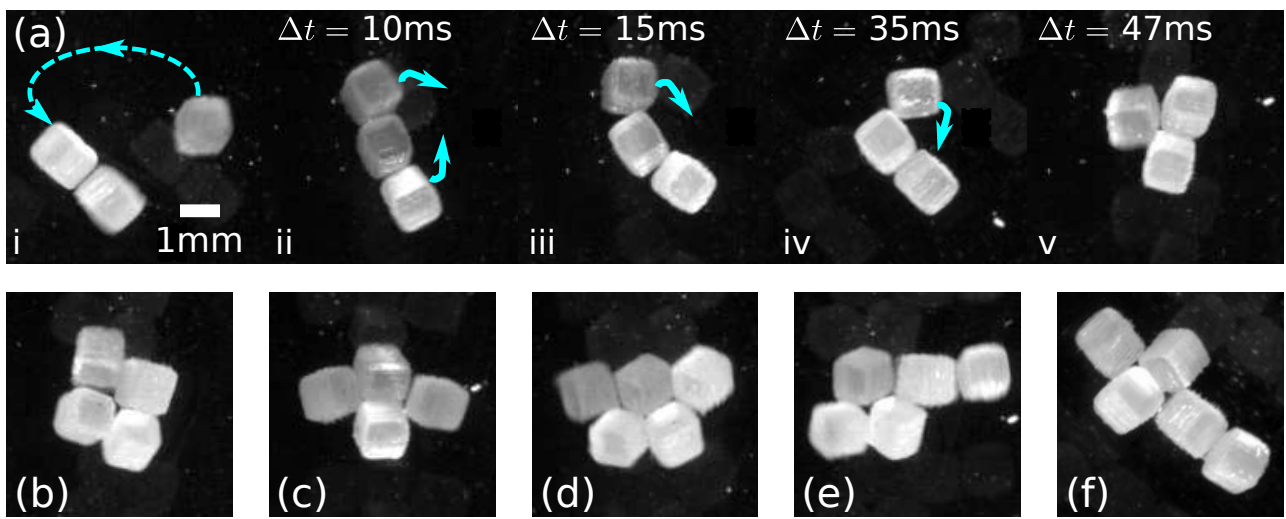}
\caption{Cluster formation driven by particle edges. (a) Sequence of images from below showing the assembly and subsequent folding of a 3-cube cluster. Note that the 3-cluster exhibits two possible configurations: a linear chain (ii) and a triangular cluster (v). See Supplementary Movies for dynamics. (b,c) Two different possible configurations of the 4-cube cluster. (d-f) Three possible configurations of the 5-cube cluster. Note that (b) and (d) correspond most closely to the isostatic cluster shapes of four and five spheres respectively}
\label{fig:clusters}     
\end{figure*}

As an example, Fig.~\ref{fig:clusters}(a) shows the assembly of two possible configurations of the 3-cube cluster: a linear chain of cubes (ii), and a close-packed configuration where three cubes share a corner (v). Only one of these shapes (close-packed) is available to a cluster of three spheres, since a linear chain of three spheres does not have enough bonds to be rigid, \edits{and will thus fold into a close-packed arrangement}. In contrast, \edits{the edge-edge contacts of the cubes have finite extent, so that local contact forces along the length of the contact can stabilise linear chains of cubes}. \edits{These linear chains} also satisfy the condition that all particles in the cluster only interact via their edges. 

When active fluctuations in the acoustic field become strong enough \edits{(the acoustic trap is far from resonance), the clusters fluctuate up and down in the trap, occasionally colliding with the reflector. These collisions can break bonds between particles in a cluster, disrupting the linear configuration of cubes.} The cluster then folds into its other, close-packed configuration (Fig.~\ref{fig:clusters}(a), ii-v). During this process, one of the cubes detaches partially from the rest of the cluster, leaving one of its corners attached. It then pivots about this corner to a new position on the cluster, ultimately joining the other two cubes at the connecting edge between them. Despite the fact that the close-packed 3-cube cluster forces a contact between an edge and a face, we observe that it is stable to further fluctuations in the acoustic field, which change the angle of each ``hinge'' in the cluster but do not trigger dramatic rearrangements. 

These two structural motifs -- linear and close-packed -- form the basis of all larger cluster shapes. Fig.~\ref{fig:clusters}(b) and (c) shows two of the possible compact four-cube clusters (the linear chain of four cubes is not shown). The cluster shown in Fig.~\ref{fig:clusters}(c) was formed from a close-packed group of three cubes, with an additional cube added to \edits{the side of the (triangular)} cluster. \edits{In contrast, the cluster shown in Fig.~\ref{fig:clusters}(b) was formed from a pair of two-cube clusters that adjoined side on, and thus has a different set of internal angles and symmetries}. Although the two clusters look similar, since the internal angles of each cluster are incommensurate with the other, rearranging the constituent particles between the shapes requires extensive reconfiguration through the close-packed three-cube arrangement.  

Adding another cube dramatically increases the number of available configurations, only three of which we show here (Fig.~\ref{fig:clusters}(d), (e), and (f)). Each configuration achieves structural stability by mixing linear chains and close-packed motifs. For instance, the configuration shown in Fig.~\ref{fig:clusters}(d) resembles the isostatic arrangement of five spheres in two dimensions, and consists of a pair of linear chains arranged side by side. Alternatively, the same cluster can be read as a pair of close-packed 3-cube clusters that share a cube. We suggest that this configuration may be interpreted as the ``ground-state" for the 5-cluster, because it involves the fewest number of edge-face contacts. However, the degree of cooperativity required to form this structure suggests that five cubes in an acoustic trap are unlikely to generically form this cluster shape without some degree of rearrangement. 

Figures~\ref{fig:clusters}(e) and (f) show alternative 5-cube clusters, each of which is formed from a close-packed 3-cube cluster, one site of which is attached to a linear chain. We observe that adding a cube onto the 4-cluster in (b) results in the 5-cluster in (e), while adding a cube to the cluster in (c) produces the cluster in (f). Importantly, rearranging (e) into the ``ground-state" configuration in (d) requires a single cube pivoting into position. In contrast, because the 4-cluster in (c) is geometrically incommensurate with the packing that leads to the ground state of the 5-cluster system, rearranging (f) into (d) requires the deconstruction of the cluster into the appropriate 3-cube motifs. 

\section{Discussion}

We have presented acoustic levitation as a novel means to tune anisotropic, attractive interactions in a granular material. We show that particles in an acoustic field are strongly attracted to sharp edges, leading to the formation of edge-edge (rather than face-face or face-edge) contacts between cubes, cylinders, and tetrahedra. These contacts support hingelike motions, which result from a combination of the confining acoustic field and short-range forces from acoustic scattering. Simulations show that the magnitude of these short-range forces is controlled by the radius of curvature of the particle edges. \edits{Our results apply to a wide range of assembly environments (any medium that supports an acoustic standing wave) and materials (anything that scatters sound), opening possibilities for the assembly of complex granular structures from directed interactions. Future work could also utilise these directional acoustic forces to design and 3D print particles with shapes that optimise their packing, or their dynamics, in an acoustic trap.}

The interplay of contact forces due to particle shape and edge-edge interactions due to acoustic forces produces a diverse range of cluster shapes. The structure and dynamics of these clusters, particularly in the large number limit, raises interesting questions about geometric frustration in self-assembly, ergodicity in activated granular matter, and the effect of attractive forces on jamming.

\begin{acknowledgements}
We thank A.~Souslov, L.~Roth, A.~Kline, V.~Vitelli, and T.~Witten for insightful discussions. This work was supported by the National Science Foundation through grants DMR-1309611 and DMR-1810390.
\end{acknowledgements}

\bibliographystyle{spphys}

\begin{thebibliography}{10}
\providecommand{\url}[1]{{#1}}
\providecommand{\urlprefix}{URL }
\expandafter\ifx\csname urlstyle\endcsname\relax
  \providecommand{\doi}[1]{DOI \discretionary{}{}{}#1}\else
  \providecommand{\doi}{DOI \discretionary{}{}{}\begingroup
  \urlstyle{rm}\Url}\fi

\bibitem{Donev2004}
A.~Donev, I.~Cisse, D.~Sachs, E.A. Variano, F.H. Stillinger, R.~Connelly,
  S.~Torquato, P.M. Chaikin, Science \textbf{303}(5660), 990 (2004)

\bibitem{Degraaf2011}
J.~de~Graaf, R.~van Roij, M.~Dijkstra, Physical Review Letters
  \textbf{107}(15), 155501 (2011)

\bibitem{Roth2016}
L.K. Roth, H.M. Jaeger, Soft Matter \textbf{12}(4), 1107 (2016)

\bibitem{Haji2009}
A.~Haji-Akbari, M.~Engel, A.S. Keys, X.~Zheng, R.G. Petschek,
  P.~Palffy-Muhoray, S.C. Glotzer, Nature \textbf{462}(7274), 773 (2009)

\bibitem{Damasceno2012}
P.F. Damasceno, M.~Engel, S.C. Glotzer, Science \textbf{337}(6093), 453 (2012)

\bibitem{Neudecker2013}
M.~Neudecker, S.~Ulrich, S.~Herminghaus, M.~Schr{\"o}ter, Physical Review
  Letters \textbf{111}(2), 028001 (2013)

\bibitem{Behringer2014}
R.~Behringer, D.~Bi, B.~Chakraborty, A.~Clark, J.~Dijksman, J.~Ren, J.~Zhang,
  Journal of Statistical Mechanics: Theory and Experiment \textbf{2014}(6),
  P06004 (2014)

\bibitem{Walsh2016}
L.~Walsh, N.~Menon, Journal of Statistical Mechanics: Theory and Experiment
  \textbf{2016}(8), 083302 (2016)

\bibitem{Miszta2011}
K.~Miszta, J.~De~Graaf, G.~Bertoni, D.~Dorfs, R.~Brescia, S.~Marras,
  L.~Ceseracciu, R.~Cingolani, R.~Van~Roij, M.~Dijkstra, et~al., Nature
  Materials \textbf{10}(11), 872 (2011)

\bibitem{Azema2012}
E.~Az{\'e}ma, F.~Radjai, Physical Review E \textbf{85}(3), 031303 (2012)

\bibitem{Miskin2013}
M.Z. Miskin, H.M. Jaeger, Nature Materials \textbf{12}(4), 326 (2013)

\bibitem{Farhadi2014}
S.~Farhadi, R.P. Behringer, Physical Review Letters \textbf{112}(14), 148301
  (2014)

\bibitem{Zhao2016}
Y.~Zhao, K.~Liu, M.~Zheng, J.~Bar{\'e}s, K.~Dierichs, A.~Menges, R.P.
  Behringer, Granular Matter \textbf{18}(2), 24 (2016)

\bibitem{Tang2016}
J.~Tang, R.P. Behringer, Europhysics Letters \textbf{114}(3), 34002 (2016)

\bibitem{Murphy2018}
K.A. Murphy, K.A. Dahmen, H.M. Jaeger, Physical Review X \textbf{9}(1), 011014
  (2019)

\bibitem{Cox2016}
M.~Cox, D.~Wang, J.~Bar{\'e}s, R.P. Behringer, Europhysics Letters
  \textbf{115}(6), 64003 (2016)

\bibitem{Royer2009}
J.R. Royer, D.J. Evans, L.~Oyarte, Q.~Guo, E.~Kapit, M.E. M{\"o}bius, S.R.
  Waitukaitis, H.M. Jaeger, Nature \textbf{459}(7250), 1110 (2009)

\bibitem{Lee2015}
V.~Lee, S.R. Waitukaitis, M.Z. Miskin, H.M. Jaeger, Nature Physics
  \textbf{11}(9), 733 (2015)

\bibitem{Lumay2008}
G.~Lumay, N.~Vandewalle, Physical Review E \textbf{78}(6), 061302 (2008)

\bibitem{Cavallaro2011}
M.~Cavallaro, L.~Botto, E.P. Lewandowski, M.~Wang, K.J. Stebe, Proceedings of
  the National Academy of Sciences \textbf{108}(52), 20923 (2011)

\bibitem{Gorkov1961}
L.~Gorkov, Sov. Phys. Doklady \textbf{6}, 773 (1962)

\bibitem{Doinikov1994}
A.~Doinikov, Proceedings of the Royal Society of London. Series A: Mathematical
  and Physical Sciences \textbf{447}(1931), 447 (1994)

\bibitem{Bruus2012}
H.~Bruus, Lab on a Chip \textbf{12}(6), 1014 (2012)

\bibitem{Settnes2012}
M.~Settnes, H.~Bruus, Physical Review E \textbf{85}(1), 016327 (2012)

\bibitem{Silva2014}
G.T. Silva, H.~Bruus, Physical Review E \textbf{90}(6), 063007 (2014)

\bibitem{Wei2004}
W.~Wei, D.B. Thiessen, P.L. Marston, The Journal of the Acoustical Society of
  America \textbf{116}(1), 201 (2004)

\bibitem{Wang2011}
J.~Wang, J.~Dual, The Journal of the Acoustical Society of America
  \textbf{129}(6), 3490 (2011)

\bibitem{Mitri2016}
F.~Mitri, Physics of Fluids \textbf{28}(7), 077104 (2016)

\bibitem{Crum1971}
L.A. Crum, The Journal of the Acoustical Society of America \textbf{50}(1B),
  157 (1971)

\bibitem{Leighton1990}
T.~Leighton, A.~Walton, M.~Pickworth, European Journal of Physics
  \textbf{11}(1), 47 (1990)

\bibitem{Westervelt1951}
P.J. Westervelt, The Journal of the Acoustical Society of America
  \textbf{23}(3), 312 (1951)

\bibitem{Doinikov1997}
A.A. Doinikov, The Journal of the Acoustical Society of America
  \textbf{101}(2), 713 (1997)

\bibitem{Glynne2013}
P.~Glynne-Jones, P.P. Mishra, R.J. Boltryk, M.~Hill, The Journal of the
  Acoustical Society of America \textbf{133}(4), 1885 (2013)

\bibitem{Hahn2015}
P.~Hahn, I.~Leibacher, T.~Baasch, J.~Dual, Lab on a Chip \textbf{15}(22), 4302
  (2015)

\bibitem{Leibacher2015}
I.~Leibacher, P.~Hahn, J.~Dual, Microfluidics and Nanofluidics \textbf{19}(4),
  923 (2015)

\bibitem{Rudnick1990}
J.~Rudnick, M.~Barmatz, The Journal of the Acoustical Society of America
  \textbf{87}(1), 81 (1990)

\bibitem{Lim2018}
M.X. Lim, A.~Souslov, V.~Vitelli, H.M. Jaeger, Nature Physics \textbf{15}(5),
  460 (2019)

\bibitem{Andrade10}
M.A.B. Andrade, F.~Buiochi, J.C. Adamowski, Ultrasonics, Ferroelectrics, and
  Frequency Control, IEEE Transactions on, \textbf{57}(2), 469:479 (2010)

\bibitem{Garcia2014}
A.~Garcia-Sabat{\'e}, A.~Castro, M.~Hoyos, R.~Gonz{\'a}lez-Cinca, The Journal
  of the Acoustical Society of America \textbf{135}(3), 1056 (2014)

\bibitem{Mohapatra2018}
A.R. Mohapatra, S.~Sepehrirahnama, K.M. Lim, Physical Review E \textbf{97}(5),
  053105 (2018)

\bibitem{Vutukuri2014}
H.R. Vutukuri, F.~Smallenburg, S.~Badaire, A.~Imhof, M.~Dijkstra,
  A.~Van~Blaaderen, Soft Matter \textbf{10}(45), 9110 (2014)

\bibitem{Rossi2011}
L.~Rossi, S.~Sacanna, W.T. Irvine, P.M. Chaikin, D.J. Pine, A.P. Philipse, Soft
  Matter \textbf{7}(9), 4139 (2011)

\bibitem{Zhao2011}
K.~Zhao, R.~Bruinsma, T.G. Mason, Proceedings of the National Academy of
  Sciences \textbf{108}(7), 2684 (2011)

\bibitem{vanAnders2013}
G.~van Anders, N.K. Ahmed, R.~Smith, M.~Engel, S.C. Glotzer, ACS Nano
  \textbf{8}(1), 931 (2013)

\bibitem{VanAnders2014}
G.~van Anders, D.~Klotsa, N.K. Ahmed, M.~Engel, S.C. Glotzer, Proceedings of
  the National Academy of Sciences \textbf{111}(45), E4812 (2014)

\end{thebibliography}

\end{document}